\documentclass[aps,pre,twocolumn,groupedaddress,showpacs]{revtex4-1}
\usepackage{graphicx}
\usepackage{dcolumn}
\usepackage{bm}
\usepackage{amsmath}
\usepackage{float}
\usepackage{subfigure}
\begin{document}

%\preprint{}

\title{Principle of Minimal Work Fluctuations}


\author{Gaoyang Xiao and Jiangbin Gong}
\email[]{phygj@nus.edu.sg}
%\thanks{}
\affiliation{Department of Physics and Centre for Computational Science and Engineering, National University of Singapore, Singapore 117542}
%\affiliation{$^2$Department of Physics and Centre for Computational Science and Engineering, National University of Singapore, Singapore 117542}


\date{\today}

\begin{abstract}
Understanding and manipulating work fluctuations in microscale and nanoscale systems are of both fundamental and practical interest. For example, in considering the Jarzynski equality $\langle e^{-\beta W} \rangle=e^{-\beta \Delta F}$, a change in the fluctuations of $e^{-\beta W}$ may impact on how fast the statistical average of $e^{-\beta W}$ converges towards the theoretical value $e^{-\beta \Delta F}$, where $W$ is the work, $\beta$ is the inverse temperature, and $\Delta F$ is free energy difference between two equilibrium states. Motivated by our previous study aiming at the suppression of work fluctuations, here we obtain a principle of minimal work fluctuations. In brief, adiabatic processes as treated in quantum and classical adiabatic theorems yield the minimal fluctuations in $e^{-\beta W}$. In the quantum domain, if a system initially prepared at thermal equilibrium is subject to a work protocol but isolated from a bath during the time evolution, then a quantum adiabatic process without energy level crossing (or an assisted adiabatic process reaching the same final states as in a conventional adiabatic process) yields the minimal fluctuations in $e^{-\beta W}$, where $W$ is the quantum work defined by two energy measurements in the beginning and at the end of the process. In the classical domain where the classical work protocol is realizable by an adiabatic process, then the classical adiabatic process also yields the minimal fluctuations in $e^{-\beta W}$.  Numerical experiments based on a Landau-Zener process confirm our theory in the quantum domain, and our theory in the classical domain explains our previous numerical findings regarding the suppression of classical work fluctuations [G.~Y.~Xiao and J.~B.~Gong, Phys. Rev. E {\bf 90}, 052132 (2014)].
\end{abstract}

%\keywords{}

\maketitle

\section{Introduction}
In small systems involving few degrees of freedom, thermal fluctuations and quantum fluctuations in work, heat, and other quantities can be comparable to their ensemble mean values.  It is of fundamental interest to understand and control these fluctuations.  To that end, fluctuation theorems (e.g., the Jarzynski equality \cite{Jarzynski.97.PRL,Jarzynski.97.PRE,Mukamel.03.PRL,Tasaki.00.apc,hanggireview} and the Crooks theorem \cite{Crooks.99.PRE}) constitute foundational results because they offer rigorous relations between nonequilibrium statistical fluctuations with equilibrium properties.
Remarkably, the existence of these fluctuation theorems does not rule out the possibility to further
 manipulate the fluctuations in work and heat. This was clearly shown, for example, in Ref.~\cite{Xiao.14.PRE}.
 As indicated by the Jarzynski equality, i.e., $\langle e^{-\beta W} \rangle=e^{-\beta \Delta F}$ (where $W$ is the work done on a system initially prepared at thermal equilibrium with inverse temperature $\beta$, $\Delta F$ is the free energy difference between the initial and final equilibrium states of the same $\beta$), the mean value of $e^{-\beta W}$ is determined by $\beta$ and $\Delta F$, but the fluctuations in the work statistics, as manifested in the variance of $e^{-\beta W}$,  can still be suppressed by an additional control field \cite{Xiao.14.PRE, Deng.13.PRE}. Such suppression in the work fluctuations may be beneficial in achieving a better convergence of the statistical average of $e^{-\beta W}$ towards the theoretical value $e^{-\beta \Delta F}$.  This hints the possibility in boosting the performace of the Jarzynski equality in an actual application.

There is a second reason for us to focus on the work fluctuations.  In particular,
advances in our knowledge of thermodynamics applied to small systems have simulated studies of efficient energy devices (classical or quantum) at the microscale and nanoscale \cite{Abah.12.PRL,Bergenfeldt.14.PRL,Zhang.14.PRL}. In these devices the work fluctuations are shown to be significant and the suppression of the work fluctuations is seen to be related to the heat-to-work conversion efficiency  \cite{Deng.13.PRE,Zheng.14.PRE,Zheng.15.a}.  Indeed, our recent study \cite{gaoyang-preprint} also showed that the heat-to-work efficiency in nanoscale Carnot cycles may not reach the Carnot efficiency because,  in general, such energy devices are operating at nonequilibrium conditions and understanding the effect of work fluctuations therein is of fundamental interest.
A natural question then follows.  Is there a lower bound of the work fluctuations and if yes,  how  to reach the lower bound?

Here we restrict ourselves to work protocols in thermally isolated systems, i.e., in the absence of a heat bath.
These processes can be one important step in the cyclic operations of a quantum heat engine, such as in Carnot cycles and Otto cycles \cite{gaoyang-preprint,Bender.00.JPAMG,Abah.12.PRL,Quan.07.PRE}.
The energy change of the system in these processes just reflects the work done to the system.  This paper considers the quantum domain first, where the quantum work is defined via two energy measurements in the beginning and at the end of a process.  In particular, given a quantum state initially prepared at thermal equilibrium with inverse temperature $\beta$ (the same initial state preparation was assumed in deriving the Jarzynski equality), we shall demonstrate that a quantum (mechanical) adiabatic process (as defined in the celebrated quantum adiabatic theorem \cite{adiabatic}), or a corresponding controlled process reaching the same final states as in the conventional adiabatic process, yields the minimal fluctuations in an exponential form of the quantum work, namely, $e^{-\beta W}$.  The implicit assumption for this principle to apply
is that the ordering of the energy levels of the initial and final Hamiltonians does not change.
Thus, compared with any other unitary evolution to realize a work protocol in a thermally isolated system, a quantum (mechanical) adiabatic process, executed slowly or executed fast via a control field,
achieves a lower bound in the fluctuations of $e^{-\beta W}$.  We term this original finding as the {\it principle of minimal work fluctuations}.  Numerical experiments based on a Landau-Zener model \cite{LZ1,LZ2} confirm our theory.

The principle of minimal quantum work fluctuations is also extended to the classical domain.  The motivation of this extension is two-fold.  First, the proof of the principle in the quantum domain clearly indicates the existence of a semiclassical analog. In particular, the quantum energy level index $n$, which is important in our quantum proof, reminds us of the classical action variable $I$ because in semiclassical quantization,  $n$ and $I$ are simply related.  Second, in our previous study aiming at the suppression of work fluctuations in a classical system, we observed that an optimal control field can dramatically suppress the fluctuations in $e^{-\beta W}$ but can never outperform that achieved from a classical (mechanical) adiabatic process in all the cases studied (within statistical error) \cite{Xiao.14.PRE}. Indeed, we are able to explicitly show, at least for one-dimensional classical systems, that a classical adiabatic process yields the minimal fluctuations in $e^{-\beta W}$.

This paper is arranged as follows. In Sec.~II, we review the definition of quantum work with necessary details. In Sec.~III, we  calculate the fluctuations in $e^{-\beta W}$ and give a detailed proof of the principle of minimal fluctuations in quantum  work. This theory is tested in Sec.~IV in a finite-time Landau-Zener process, by comparing the fluctuations attained using an optimal control approach with those obtained from assisted adiabatic passage. In Sec.~V, we extend our minimal fluctuation principle to the classical domain.
%%%%%%%%%%%%%%%%%%%%%%%%%%%%%%%%%%%%%%%%%%%%%%%%%%%%%%%%%%%%%%%%%%

\section{Quantum work and related quantities based on two-time energy measurements}
Let us consider a quantum system with a Hilbert space of dimension $N$ ($N$ can be infinity).  The system is subject to a unitary process described by a time-dependent Hamiltonian $H(t)$ starting from $t=0$ to $t=\tau$. For $i=1,\ldots,N$, let $|\psi_i\rangle$ and $|\psi^\prime_i\rangle$ denote the eigenstates of $H_0\equiv H(0)$ and $H_\tau\equiv H(\tau)$, respectively, with eigenvalues $E_i$ and $E^\prime_i$. Let $U$ denote the unitary time evolution operator associated with the whole process from $t=0$ to $t=\tau$.

At $t=0$, the system is assumed to be prepared in the Gibbs state with inverse temperature $\beta$, with its equilibrium density matrix given by $\rho_0$:
\begin{equation}\label{density1}
  \rho_0=\sum_{i=1}^{N}p_i|\psi_i\rangle \langle \psi_i|,
\end{equation}
with
\begin{equation}\label{Pi1}
  p_i=\frac{e^{-\beta E_i}}{\sum_{i=1}^N e^{-\beta E_i}}\equiv \frac{1}{Z^q_0}e^{-\beta E_i},
\end{equation}
where $Z_0^q$ is the quantum partition function.

At the end of the work protocol, one asks how much work has been done to the system. This can be answered by considering two energy measurements at $t=0$ and $t=\tau$ \cite{Tasaki.00.apc,hanggireview}.
Indeed, for an arbitrary  function $f(E,E^\prime)$ involving two energies  $E$ and $E^\prime$ associated with $H_0$ and $H_\tau$ , one needs to perform such kind of two-time measurements, with the average value of $f(E,E^\prime)$ given by
\begin{equation}\label{average}
  \overline{f(E,E^\prime)}=\sum_{i,j=1}^{N}p_i|a_{ij}|^2f(E_i,E^\prime_j),
\end{equation}
where
\begin{equation}\label{transition}
  a_{ij}=\langle \psi^\prime_j| U |\psi_i \rangle,
\end{equation}
with
\begin{equation}\label{normalize}
  \sum_{i=1}^N |a_{ij}|^2=\sum_{j=1}^N |a_{ij}|^2=1.
\end{equation}
Experimentally, $p_i|a_{ij}|^2$ can be interpreted as the probability that one finds the system in the $i$-th eigenstate of $H_0$ at $t=0$ and then in the $j$-th eigenstate of $H_\tau$ at $t=\tau$.

Following this definition based on two-time energy measurements, the average work done on an isolated system is given by
\begin{equation}\label{work1}
  \begin{split}
    \langle W \rangle & =\overline{E-E^\prime} = \sum_{i,j=1}^n p_i |a_{ij}|^2(E_j^\prime-E_i)\\
      & = \text{Trace}[U \bar{\rho}_0 U^\dag H_\tau]-\text{Trace}[\bar{\rho}_0 H_0],
  \end{split}
\end{equation}
where $\bar{\rho}_0$ is the density matrix right after the first energy measurement at $t=0$. For our case or any other initial state as a completely mixed state in terms of energy eigenstates, $\bar{\rho}_0=\rho_0$.

We note in passing that with the above-mentioned initial equilibrium state, another definition of the quantum work, i.e., $d\langle W \rangle = \text{Trace} [dH \rho]$ \cite{Gelbwaser-Klimovsky.15.a} yields the same expression. That is,
\begin{equation}\label{work2}
\begin{split}
  \langle W \rangle & = \int_{0}^{\tau} \text{Trace} [dH(t) \rho(t)]\ dt \\
    & = \int_{0}^{\tau}d \{\text{Trace}[H(t)\rho(t)]\} \\
    & = \text{Trace}[H_\tau \rho_\tau]-\text{Trace}[H_0 \rho_0],
\end{split}
\end{equation}
where $\rho(t)$ is the time evolving density matrix and we have used $\text{Trace}[d\rho(t)H(t)]=0$ for a thermally isolated system under unitary evolution.
%%%%%%%%%%%%%%%%%%%%%%%%%%%%%%%%%%%%%%%%%%%%%%%%%%%%%%%%%%%%%%%%%%%%%%%%%%%%%%%%%%%%%%
\section{Principle of minimal work fluctuations: quantum systems}
In the literature, there is the minimal work principle \cite{Allahverdyan.05.PRE} that has been attracting considerable attention recently.  That is, for an isolated system initially prepared as a Gibbs equilibrium distribution described above
[Eq.~(\ref{density1})], a quantum (mechanical) adiabatic process, if implementable, yields a minimized average work [see Eq.~(\ref{work1})] for the system among all possible thermally isolated processes that start from $H_0$ and end with $H_\tau$.
Of particular interest to studies of nanoscale energy devices, this minimal work principle implies that, in order to minimize the so-called disspated work and maximize the heat-to-work conversion efficiency of a quantum heat engine, a thermally isolated step   should implement bare adiabatic processes or accelerated adiabatic processes \cite{gaoyang-preprint}.  Both as a fundamental question and a practical issue (relevant to understanding  the stability or reliability of the work output of nanoscale energy devices), here we hope to understand the lower bound in the work fluctuations.  A digestion of the key element in the proof of the minimal work principle, together with the Jarzynski equality,  stimulates us to consider fluctuations in $e^{-\beta W}$, rather than in $W$ directly.

For convenience, when referring to a general adiabatic process, we mean both a conventional adiabatic process or its alternative version based on engineered time-dependent fields, such as those processes named ``shortcuts to adiabaticity" (STA) \cite{Rice,Berry,Torrontegui.13.AAMOP,Chen.10.PRL,Ibanez.12.PRL,Campo.13.PRL}.
We order the initial eigenenergy values as
\begin{equation}\label{spectrum1}
  E_1 < E_2<  \cdots < E_N.
\end{equation}
It follows that the thermal excitation probabilities have the following ordering
\begin{equation}\label{orderp1}
  p_1 > p_2>\cdots > p_N.
\end{equation}
Assuming that $H_0$ and $H_\tau$ can be connected by a quantum adiabatic process, then we require the ordering of $E_i'$ (eigenvalues of $H_\tau$) is the same as $E_i$, namely,
\begin{equation}\label{spectrum1}
  E'_1 < E'_2 < \cdots < E'_N.
\end{equation}
This leads to
\begin{equation}
e^{-2\beta E'_{j+1}} < e^{-2\beta E'_{j}}, \ j=1,2, \cdots, N-1.
\end{equation}
which will be useful below.

We are now ready to prove that a quantum adiabatic process  produces a minimized variance in
 $e^{-\beta W}$, as compared with all other thermally isolated processes that start from $H_0$ and end with $H_\tau$.
To distinguish from work in other cases,
we use $\widetilde{W}$ to denote the work in a quantum adiabatic process and use $\widetilde{a}_{ij}$ to represent the associated transition probabilities from state $E_i$ to state $E'_j$ [see Eq.~(\ref{transition})].

The fluctuations of $e^{-\beta W}$, characterized by the variance of $e^{-\beta W}$, can be quantified by
\begin{equation}\label{variance}
  \begin{split}
    \sigma ^2 \left(e^{-\beta W}\right )& = \left\langle \left(e^{-\beta W}-\langle e^{-\beta W}\rangle\right)^2 \right\rangle\\
      & = \left\langle e^{-2 \beta W}\right\rangle - \left\langle e^{-\beta W}\right\rangle ^2.
  \end{split}
\end{equation}
According to the Jarzynski equality $\langle e^{-\beta W} \rangle = e^{-\beta \Delta F}$, the second term of the right-hand side of
 Eq.~(\ref{variance}) is a constant for fixed $H_0$ and $H_\tau$. Thus, essentially we are left with treating the first term only. According to  Eq.~(\ref{average}), the first term can be evaluated by
\begin{equation}\label{variance1}
  \langle e^{-2 \beta W} \rangle = \sum_{i,j=1}^N p_i |a_{ij}|^2 e^{-2 \beta (E^\prime_j-E_i)}.
\end{equation}
Using the identity (summation by parts)
\begin{equation}\label{Sbypart}
  \sum_{j=1}^N b_j c_j = b_N \sum_{j=1}^N c_j - \sum_{j=1}^{N-1}\left(b_{j+1}-b_j\right) \sum_{k=1}^j c_k,
\end{equation}
we obtain from Eq.~(\ref{variance1}) and Eq.~(\ref{normalize}) the following:
\begin{widetext}
\begin{equation}\label{variance2}
\begin{split}
  \langle e^{-2 \beta W} \rangle & = \sum_{i=1}^N p_i e^{2 \beta E_i} \left[\sum_{j=1}^N e^{-2 \beta E_N^\prime} |a_{ij}|^2-\sum_{j=1}^{N-1}\left(e^{-2\beta E_{j+1}^\prime}-e^{-2\beta E_j^\prime}\right)\sum_{k=1}^j|a_{ik}|^2\right] \\
    & = e^{-2\beta E_N'} \sum_{i=1}^N p_i e^{2 \beta E_i} - \sum_{j=1}^{N-1}\left(e^{-2\beta E_{j+1}^\prime}-e^{-2\beta  E_j^\prime}\right)\sum_{i=1}^N p_i e^{2 \beta E_i} \sum_{k=1}^{j}|a_{ik}|^2.
\end{split}
\end{equation}
%\end{widetext}
In the same fashion, one can look into the variance in $e^{-\beta \widetilde{W}}$, which is determined by
the ensemble average $\langle e^{-2\beta \widetilde{W}}\rangle$, with
%\begin{widetext}
\begin{equation}\label{variance3}
\begin{split}
  \langle e^{-2 \beta \widetilde{W}} \rangle & = \sum_{i=1} ^N p_i e^{2 \beta E_i} \left[\sum_{j=1}^N e^{-2 \beta E_N^\prime} |a_{ij}|^2-\sum_{j=1}^{N-1}\left(e^{-2\beta E_{j+1}^\prime}-e^{-2\beta E_j^\prime}\right)\sum_{k=1}^j|\widetilde{a}_{ik}|^2\right] \\
    & = e^{-2\beta E_N'} \sum_{i=1}^N p_i e^{2 \beta E_i} - \sum_{j=1}^{N-1}\left(e^{-2\beta E_{j+1}^\prime}-e^{-2\beta E_j^\prime}\right)\sum_{i=1}^N p_i e^{2 \beta E_i} \sum_{k=1}^{j}|\widetilde{a}_{ik}|^2.
\end{split}
\end{equation}
\end{widetext}

The difference in the fluctuations of $e^{-\beta W}$ between a general nonequilibrium process and an adiabatic process
is given by
\begin{equation}\label{difference1}
    \langle e^{-2\beta W} \rangle - \langle e^{-2\beta \widetilde{W}} \rangle =- \sum_{j=1}^{N-1}(e^{-2\beta E_{j+1}^\prime}-e^{-2\beta E_j^\prime}) \Xi_j,
\end{equation}
where
\begin{equation}\label{Xi}
  \Xi_j=\sum_{i=1}^N p_i e^{2 \beta E_i}\sum_{k=1}^{j}[|a_{ik}|^2-|\widetilde{a}_{ik}|^2].
\end{equation}
Note now that in a quantum adiabatic process, the state populations do not change, with
\begin{equation}\label{delta1}
  |\widetilde{a}_{ik}|^2=\delta_{ik}.
\end{equation}
Further using Eqs.~(\ref{density1}) and (\ref{normalize}), we obtain from Eq.~(\ref{Xi})
\begin{equation}\label{Xi1}
  \begin{split}
    \Xi_j & =\frac{1}{Z_0^q} \sum_{i=1}^N e^{\beta E_i} \sum_{k=1}^j \left[|a_{ik}|^2-|\tilde{a}_{ik}|^2\right]\\
      & =\frac{1}{Z_0^q}\sum_{i=1}^j e^{\beta E_i}\left[\sum_{k=1}^j |a_{ik}|^2-1 \right]+\frac{1}{Z_0^q}\sum_{i=j+1}^{N}e^{\beta E_i}\sum_{k=1}^{j}|a_{ik}|^2 \\
      & \geq \frac{e^{\beta E_j}}{Z_0^q} \sum_{i=1}^{j}\left[\sum_{k=1}^j |a_{ik}|^2-1\right ]+\frac{e^{\beta E_j}}{Z_0^q} \sum_{i=j+1}^{N}\sum_{k=1}^j|a_{ik}|^2 \\
      & = \frac{e^{\beta E_j}}{Z_0^q} [-j+j] \\
      & = 0.
  \end{split}
\end{equation}
That is, the function $\Xi_j$ defined above cannot be negative.  Returning to Eq.~(\ref{difference1}), we finally arrive at
\begin{equation}
\langle e^{-2\beta W} \rangle - \langle e^{-2\beta \widetilde{W}} \rangle \geq 0
\end{equation}
or equivalently,
\begin{equation}
\sigma^2\left( e^{-2\beta W} \right) \geq \sigma^2 \left( e^{-2\beta \widetilde{W}} \right).
\end{equation}
That is, the fluctuations in $e^{-\beta W}$, as characterized by the square variance of  $e^{-\beta W}$, become minimal if the process
from $H_0$ to $H_\tau$ is a quantum adiabatic process.

\begin{table*}[htb]\renewcommand{\arraystretch}{1.5}
\caption{
Performance of work fluctuation suppression in a Landau-Zener process, in the absence or presence of control fields needed for realizing STA or OCT, characterized by the variance in $e^{-\beta W}$, using $10^6$ two-time energy measurement results.
The model Hamiltonian is described in (Eq.~\ref{twolevel}), with $Z(t)$ given in Eq.~(\ref{Z(t)}), $Z(0)=1.0$, $Z(\tau)=3.0$, $\tau=0.0001$ (duration of the process), $X_0=2.0$, and the inverse temperature $\beta=0.1$. The numerically found transition probabilities are also presented, where $|1\rangle$ and $|1^\prime \rangle$ are the ground states of the initial and final Hamiltonian, $|2^\prime \rangle$ is the excited state of the final Hamiltonian, and $U$ denotes time evolution operator. The obtained values of $\langle e^{-\beta W} \rangle$ all agree with the theoretical value $e^{-\beta \Delta F}\approx 1.040$ obtained from the Jarzynski equality.}
\vskip 1em
\centering
\begin{tabular}{ccccc}\hline\hline
{\quad}Process{\quad} & ${\quad}\langle e^{-\beta W} \rangle{\quad}$ & ${\quad}\sigma (e^{-\beta W}){\quad}$ & ${\quad}|\langle 1^\prime |U|1\rangle|^2{\quad}$ & ${\quad} |\langle 2^\prime |U|1\rangle|^2 {\quad}$\\[0.5ex]
\hline
bare{\;}system & $1.040$ & $0.202$ & $0.9341$ & $0.0659$\\
\hline
STA{\;} & $1.040$ & $0.134$ & $1.000$ & $0$\\
\hline
Optimal{\;}control  & $1.040$ & $0.134$ & $1.000$ & $0$\\
\hline\hline
\end{tabular}
\label{performance}
\end{table*}

Together with the previously established minimal work principle, one can now conclude that given a thermally isolated system prepared initially at equilibrium and later driven by Hamiltonian $H(t)$ with specified $H_0$ at the start and $H_\tau$ in the end, then a quantum adiabatic process, if implementable, not only yields the minimal average work, but also the minimal fluctuations in $e^{-\beta W}$. This is one of the main results of this work.  Before closing this section,  a few remarks are in order.

First, it is necessary to highlight the counter-intuitive nature of our minimal work fluctuation principle.  Given that a quantum adiabatic process minimizes the mean work,  one may naively think that the corresponding ensemble-averaged value of $e^{-2\beta W}$, a key quantity in obtaining the variance of $e^{-\beta W}$, seems to be maximized.  But our proof indicates the opposite, namely, the ensemble-averaged value of $e^{-2\beta W}$ is actually minimized in a quantum adiabatic process, and so is the variance of $e^{-\beta W}$.

Second, throughout the proof, we have assumed that the ordering of $E_i'$ is the same as $E_i$ to ensure the existence of an adiabatic process to connect $H_0$ to $H_\tau$.  If this prerequisite is not satisfied, then it is unclear what kind of process can reach a lower bound of work fluctuations. Indeed, there a very slow process not necessarily yields very small work fluctuations.  By itself this constitutes a fascinating topic for future study.

Third, we have emphasized above that a quantum adiabatic process may be executed via a control field by taking advantage of STA \cite{Deng.13.PRE,Xiao.14.PRE,Berry,Rice,Torrontegui.13.AAMOP}. Consider, for example, one type of STA, which is also called counter-diabatic driving \cite{Rice} or transitionless quantum driving \cite{Berry}. There
an additional control Hamiltonian is introduced to drive a system. In this case, the work to the system is done by two sources: the original protocol to tune the system from $H_0$ to $H_\tau$, as well as the additional control field implementing an accelerated adiabatic process.  In practice, one may wish to set the additional control field achieving STA to zero at initial and final times, so that the energy measurements of the total Hamiltonian in the beginning and at the end still reflect the inherent energy eigenvalues of the bare system alone.  With that understanding in mind, it is now possible to achieve  minimal quantum work fluctuations using very fast processes!

%Of particular interest, some people regard those quantum adiabatic processes as the conventional quasi-static thermally isolated %processes since both of them are infinitely slow, while the crucial differences are listed below:
%\begin{itemize}
%  \item[(i)] Conventional quasi-static processes indicates that the system is always at equilibrium, while there is no reason to %expect $\rho(t)$ evolved by unitary operator from equilibrium initial state~(\ref{density1}) should also be equilibrium (${e^{-\beta %H(t)}}/{Z_t}$) during the quantum adiabatic processes \cite{Sekimoto.00.PRE,Allahverdyan.05.PRE}.
%  \item[(ii)] In finite systems, both minimal work principle and minimal fluctuation principle require no energy level crossing %additionally. That is, if energy levels indeed cross, a fast process may lead to smaller work and fluctuations than infinitely slow %one \cite{Allahverdyan.05.PRE}. Such exceptions do not exist for conventional quasi-static processes.
%  \item[(iii)] To our knowledge, there is no method to accelerate a quasi-static process so far. However, shortcuts to adiabaticity %have been studied to realize quantum adiabatic processes within a finite time period %\cite{Berry.09.JPAMT,Campo.13.PRL,Torrontegui.13.AAMOP,Deffner.14.PRX}. It is thus possible to construct a quantum heat engine with %high efficiency, large output power and low fluctuations simultaneously.
%\end{itemize}
%%%%%%%%%%%%%%%%%%%%%%%%%%%%%%%%%%%%%%%%%%%%%%%%%%%%%%%%%%%%%%%%%%%%%%%%%%%%%%%%%%%%%%%%%%
\section{Principle of minimal work fluctuations manifested in a Landau-Zener process}

\
\begin{figure}
  \centering
  \includegraphics[width=8.4cm]{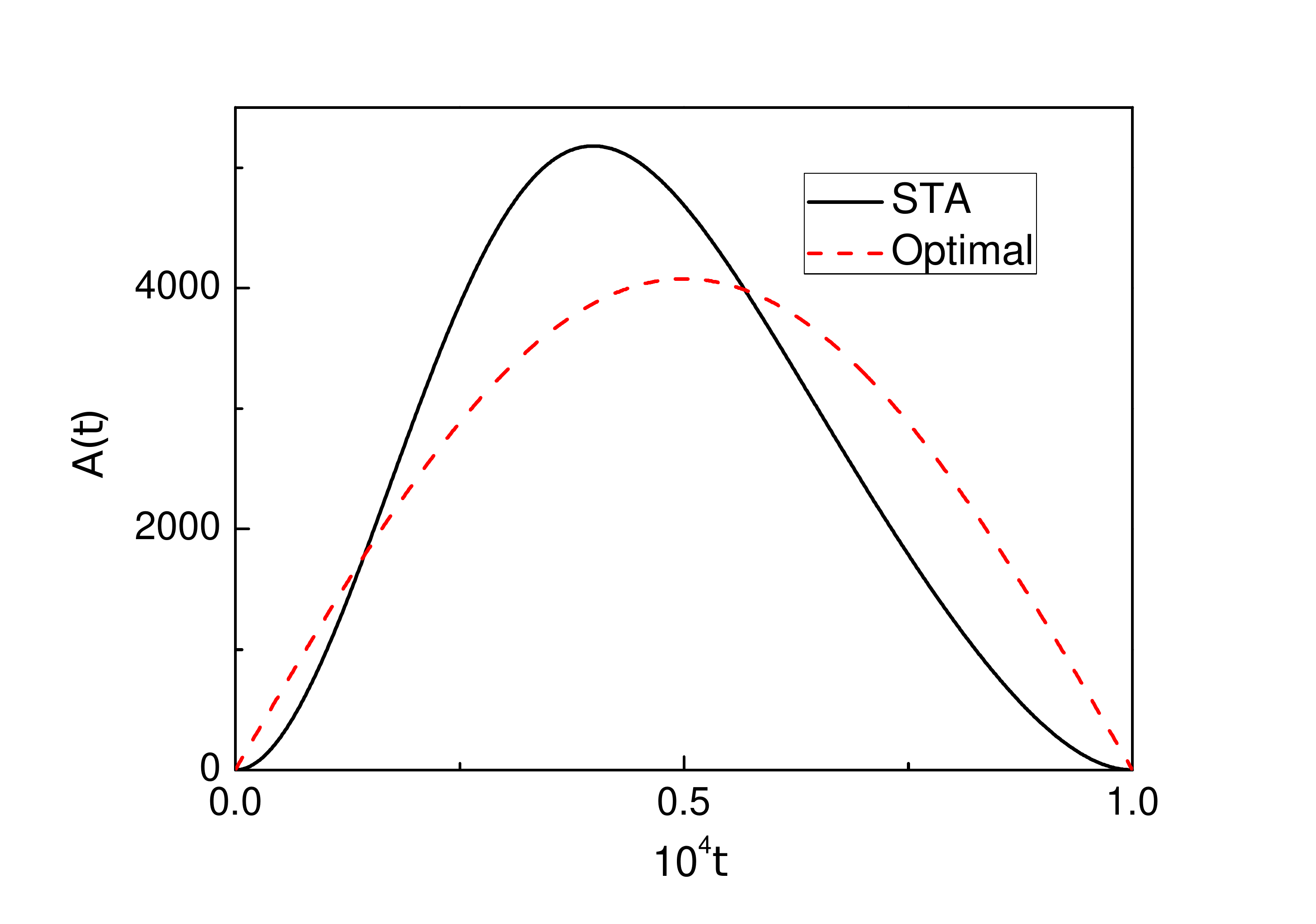}\\
  \caption{(color online) Time dependence of the  amplitude of a numerically found OCT field, as compared with that of a control field realizing STA in a two-level system (Eq.~\ref{twolevel}),  with  $Z(t)$ given in Eq.~(\ref{Z(t)}), $Z(0)=1.0$, $Z(\tau)=3.0$, $\tau=0.0001$ (duration of the process), $X_0=2.0$, and the inverse temperature set to be $\beta=0.1$. All the plotted quantities here are scaled and hence in dimensionless units.}\label{Fig}
\end{figure}

Here we use the Landau-Zener model \cite{LZ1,LZ2} to illustrate and check the principle of minimal work fluctuations.
On the one hand, we use STA to realize a fast adiabatic passage in this model and then examine the work fluctuations. On the other hand, we
use an optimal control theory (OCT) \cite{Peirce.88.PRA,Shi.90.JCP,Shi.91.CPC,Judson.92.PRL} to minimize the work fluctuations, with the variance in $e^{-\beta W}$ to be minimized. Some necessary details regarding an ensemble-based OCT approach are presented in Appendix A for completeness.   According to our principle of minimal work fluctuations, the results from our OCT can never beat that from an adiabatic process.  This fact will be numerically checked below.  Furthermore, it is curious to see how the results from OCT may approach that based on STA.

The Landau-Zener  model Hamiltonian, all in dimensionless units, is assumed to be
\begin{equation}\label{twolevel}
  H(t)=Z(t)\sigma_z+X_0\sigma_x,
\end{equation}
where $\sigma_z$ and $\sigma_x$ are the Pauli matrices.
For a time-dependent $Z(t)$ and a time-independent $X_0$, the control field achieving STA is given by \cite{Berry,Torrontegui.13.AAMOP,Chen.10.PRL,Ibanez.12.PRL}
\begin{equation}\label{STA}
  H^{\text{STA}}(t)=\hbar (\dot{\Theta}_0/2)\sigma_y,
\end{equation}
where $\Theta_0=\arccos(Z(t)/R_0)$ and $R_0=\sqrt{X_0^2+Z(t)^2}$.
As an example, we choose \cite{Campo.13.PRL}
\begin{equation}\label{Z(t)}
\begin{split}
  Z(t) & = Z(0)+[(Z(\tau)-Z(0)](\frac{t}{\tau})^3\\
    & -15[Z(\tau)-Z(0)](\frac{t}{\tau})^4+6[Z(\tau)-Z(0)](\frac{t}{\tau})^5
\end{split}
\end{equation}
such that the control field $H^{\text{STA}}(t)$ is zero at $t=0$ and at $t=\tau$.  In our numerical
calculations we choose $Z(0)=1.0$ and $Z(\tau)=3.0$ in dimensionless units. We further set $\tau$ to be as small as $0.0001$ (as compared with $\frac{1}{Z(0)}$), such that the process will be highly non-adiabatic were there no control field.

Parallel to this, the control filed in OCT is assumed to be  $A(t)\sigma_y$ with $A(0)=0$ and $A(\tau)=0$.  This boundary condition for $A(t)$ can be satisfied by introducing an appropriate time profile of a cost function (for details see Appendix).
It can be checked that the system here is fully controllable by $\sigma_y$. That is,  any unitary evolution operator can be yielded in a finite time by considering a control field of the $\sigma_y$ type \cite{Riviello.14.PRA}.

To quantitatively characterize the performance in suppressing the fluctuations in $e^{-\beta W}$ in STA and in OCT, we randomly sample initial energy eigenstates $|\psi_i \rangle$ according to the initial thermal probability distribution (the first measurement), then evolve them under the total Hamiltonian $H=H_0+H^{\text{STA}}$. Next, we again randomly sample energy eigenstates $|\psi^\prime_j\rangle$ according to its probability projected on the final state (the second measurement). Individual values of $W$ are obtained from $W_i=E^\prime_j-E_i$, and the fluctuations in $e^{-\beta W}$ are calculated from $
  \sigma(e^{-\beta W})=\sqrt{\frac{1}{M}\sum_{i=1}^M(e^{-\beta W_i}-\langle e^{-\beta W}\rangle)^2}$,
where the total number of ``trajectories" is chosen to be $M=10^6$.

%%%%%%%%%%%%%%%%%%%%%%%%%%%%%%%%%%%%%%%%%%%%%%%%%%%%%%%%%%%%%%%%%%%%%%%%%%%%

The variances in $e^{-\beta W}$ obtained in the bare system, under STA and under OCT are presented in Table \ref{performance}, along with the ensemble-average of $e^{-\beta W}$, as well as the transition probabilities between initial and final energy eigenstates. First of all, it is seen that all the three cases yield exactly the same average of $e^{-\beta W}$, consistent with the Jarzynski equality.  Secondly, both processes under STA and under OCT have suppressed the variance of $e^{-\beta W}$. But remarkably, the performance of OCT in suppressing  the variance of $e^{-\beta W}$ does not beat that of STA (producing the same variance here), thus confirming our expectation that an adiabatic process yields the lower bound for fluctuations in $e^{-\beta W}$.  An investigation of the transition probabilities also offers more insights. One observes that the transition probabilities obtained under OCT is actually the same as that obtained under STA.  That is, an OCT field aiming at minimizing work fluctuations and successfully reaching the lower bound of work fluctuations (as characterized by the variance of $e^{-\beta W}$) tends to reproduce, at $t=\tau$, the initial populations on each energy eigenstate. This confirms our theory from another angle.

Lastly, in Fig.~1 we compare the  time dependence of the control field numerically found in OCT with that of the STA control field.  It is seen that $A(t)$ found in OCT is not the same as in the case of STA.  The peak field amplitude in the OCT case is smaller. This difference clearly indicates that there are different solutions when it comes to minimize the work fluctuations. In our proof above,  the crucial requirement to reach the lower bound of work fluctuations is $a_{ij}=\delta_{ij}$, i.e., populations on states $E'_i$ stay the same as the populations on states $E_i$: what happens during the process does not matter.  An adiabatic process, which keeps the populations throughout the process, then offers a lower bound in work fluctuations. However, in other processes with nonadiabatic transitions during the time evolution but still the same final-state populations as in an adiabatic process, the lower bound of work fluctuations can still be reached.

\section{Principle of minimal work fluctuations: classical systems}
In our previous study \cite{Xiao.14.PRE} treating classical systems, a classical version of STA to realize classical shortcuts to adiabaticity \cite{Deng.13.PRE} and a classical OCT were considered to suppression classical work fluctuations. It was found that
the performance of OCT, as quantified by the suppression of the variance in $e^{-\beta W}$, can at most reach that
obtained from classical STA (within statistical error).  This hints the existence of a lower bound of classical work fluctuations, which might have been reached by classical adiabatic processes.  Following the technique we used in proving the principle of minimal work fluctuations in the quantum domain, here we shall prove an analogous principle in the classical domain.

Consider a time-dependent classical system with Hamiltonian $H^c(q,p,\lambda(t))$, where $(q,p)$ represents phase space coordinates, and $\lambda(t)$ is a time-dependent parameter evolving from $t=0$ to $t=\tau$, with $H_0^c\equiv H^c(p,q,\lambda(0))$ and $H_\tau^c\equiv H^c(p,q,\lambda(\tau))$.
In the absence of a bath, the inclusive work \cite{Jarzynski.07.CRP} as a function of initial conditions $(p_0,q_0)$ is given by
\begin{equation}\label{ClassHamiltonian}
  W_\tau(q_0,p_0)=H^c_\tau[q_\tau(q_0,p_0),p_\tau(q_0,p_0),\tau]-H^c_0(q_0,p_0,0),
\end{equation}
where $[q_\tau(q_0,p_0), p_\tau(q_0,p_0)]$ are simply the time-evolving phase space coordinates staring from $(q_0,p_0)$.
Then the thermal ensemble average $\langle e^{-2\beta W} \rangle$ needed for calculating the variance of $e^{-\beta W}$ is given by
\begin{equation}\label{Cvariance}
\begin{split}
   \langle e^{-2\beta W} \rangle = & \int \int \int\int dq_0 dp_0 dq dp\ e^{-2\beta [H^c_\tau(q,p)-H^c_0(q_0,p_0)]}  \\
    & \rho_0(q_0,p_0)\, \delta(q-q_\tau)\,\delta(p-p_\tau)\
\end{split}
\end{equation}
where $\rho_0(q_0,p_0)$ is the initial Gibbs distribution with inverse temperature $\beta$:
\begin{equation}\label{Cdensity}
  \rho_0(q_0,p_0) = \frac{1}{Z_0^c}e^{-\beta H^c_0(q_0,p_0)},
\end{equation}
with
\begin{equation}\label{Cpartition}
  Z_0^c=\int e^{-\beta H^c_0(q_0,p_0)} dq_0 dp_0.
\end{equation}

Next, to have a classical adiabatic process connecting $H^c_0$ to $H^c_\tau$, we further assume that $H^c(q,p,\lambda(t))$
can be written as $H^c_t(I_t)$, where $(I_t,\theta_t)$ are the action and angle variables associated with $H^c(q,p,\lambda(t))$.  That is, during a work protocol, the Hamiltonian can be always expressed as a function of its instantaneous action variables only.   In particular, at $t=0$, $H^c_0=H^c_0(I_0)$; at $t=\tau$, $H^c_\tau=H^c_\tau(I_\tau)$, where $I_0$ and $I_\tau$ are the action variables
in the beginning and at the end.  According to the classical adiabatic theorem, if the parameter $\lambda(t)$ changes slowly as compared with the internal frequency of the system ($\omega_t = \frac{\partial H^c_t(I_t)}{\partial I_t}$), then the action variable $I_t$ remains invariant during the work protocol.  Note that, as our proof below shows, what is truly essential is the invariance of the action variables at the two boundary times and an extra condition (analogous to the no-level-crossing condition in the quantum case) about $H^c_0(I_0)$ and $H^c_\tau(I_\tau)$.  That is, what actually happens for $0<t<\tau$ is not really essential in reaching the lower bound of the work fluctuations.  Note also that we have assumed that the system's Hamiltonian has only one degree of freedom.  Extending the proof below to systems with more than one degrees of freedom is possible, however the prerequisite for the principle to be applicable seems to be demanding in high-dimensional classical systems.

Equation (\ref{Cvariance}) can assume a different form if we introduce a canonical transformation from $(q,p)$ to action-angle variables $(I, \theta)$ (only as integration variables),
\begin{equation}\label{Cvariance1}
\begin{split}
   \langle e^{-2\beta W} \rangle = & \int \int \int \int dI_0 d\theta_0 dI d\theta\  e^{-2\beta [H_\tau^c(I)-H_0^c(I_0)]}\rho_0(I_0) \\
    & \delta[I_\tau(I_0,\theta_0)-I]\,\delta[\theta_\tau(I_0,\theta_0)-\theta],
\end{split}
\end{equation}
where $(I_\tau,\theta_\tau)$ is expressed as a function of the initial conditions
$(I_0,\theta_0)$. The explicit dependence of $(I_\tau,\theta_\tau)$ upon $(I_0,\theta_0)$ might not be spelled out below for
   convenience. The initial thermal density now becomes
\begin{equation}\label{Cdensity2}
  \rho_0(I_0) = \frac{1}{Z_0^c}e^{-\beta H^c_0(I_0)}.
\end{equation} For later use we also note
\begin{equation}\label{Cnormalize}
  \begin{split}
    & \int\int dI d\theta\delta[I_\tau(I_0,\theta_0)-I]\,\delta[\theta_\tau(I_0,\theta_0)-\theta]   \\
      = &\int\int dI_0 d\theta_0 \delta[I_\tau(I_0,\theta_0)-I]\,\delta[\theta_\tau(I_0,\theta_0)-\theta]\\
       = &\ 1.
  \end{split}
\end{equation}

In the quantum case we benefitted from the assumption that the ordering of the final energy eigenvalues $E'_i$ is the same as that of initial energy eigenvalues $E_i$.  This motivates
us to reorder certain integrals in terms of the value of $H_0^c$.  Specifically, we define a nonconventional type of integration as follows:
\begin{equation}\label{integralorder}
 \int f(I) [dI]_{{H}_0^c}\ \equiv \sum_{n=1}^{\infty} f[I(n)] \Delta I,
\end{equation}
where $I$ is the integration variable and $I(n+1)-I(n)=\Delta I$ represents an infinitesimal interval in the $I$ space, and
\begin{equation}\label{Cspectrum}
  H_0^c[I(n)]\leq H_0^c[I(n+1)], \ \ n=1,2 \cdots.
\end{equation}
The physical meaning of such a resummation is to order the variable $I$ according to $H_0(I)$, such that the summation
of $f[I(n)]\Delta I $ over $n$ is executed progressively according to the value of $H_0(I)$.  This makes it clear that
\begin{equation}
\int  [dI]_ {{H}_0^c}\ f(I) = \int dI\ f(I),
\end{equation}
namely, the ordering in the summation does not affect the final sum.  Because of the reordering, we also have
\begin{equation}\label{Cdensityorder}
  \rho_0[I(n)]\geq \rho_0[I(n+1)], \ \ n=1,2, \cdots.
\end{equation}
which means that the initial thermal excitation probability for $I(n)$ is no less than that for $I(n+1)$.
This fact will be useful for our proof below.

Exploiting the newly defined integral above,  Eq.~(\ref{Cvariance1}) can be rewritten as
\begin{widetext}
\begin{equation}\label{Cvariance2}
  \langle e^{-2\beta W} \rangle = \int_{I^{L}}^{I^{R}} [dI_0]_{{H}_0^c}\ e^{2\beta H_0^c}\rho_0(I_0)\int d\theta_0 \int_{I^{L}}^{I^{R}}[dI]_{{H}^c_0}\ e^{-2\beta H^c_\tau(I)} \int d\theta\ \delta [I_\tau-I] \delta [\theta_\tau-\theta ]
\end{equation}

In the above equation the lower and upper limits of the integrals over $\theta_0$ or $\theta$ are always 0 and $2\pi$.
For the reordered integral over $[dI_0]_{{H}_0^c}$ and $[dI]_{H^c_0}$, the lower and upper limits of the integrals are denoted
$I^{L}$ and $I^{R}$, namely,  the values of $I_0$ that gives rise to the lower limit and upper limit of $H_0^c(I_0)$.
Either $I^{L}$ or $I^{R}$ can be $+\infty$.   Performing integration by parts, we obtain from Eq.~(\ref{Cvariance2})
\
\begin{equation}\label{Cvariance3}
  \begin{split}
    \langle e^{-2\beta W} \rangle & = \int_{I^{L}}^{I^{R}}[dI_0]_ {{H}_0^c}\ e^{2\beta H_0^c(I_0)}\rho_0(I_0) \int d\theta_0 \int d\theta\ \delta [\theta_\tau-\theta ]  \\
    & \quad\Bigg\{ e^{-2\beta H^c_\tau(I^{R})}-\int_{I^{L}}^{I^{R}} d\left[e^{-2\beta H^c_\tau(I)}\right]_{{H}_0^c} \int_{I^{L}}^{I}[dI^\prime]_{{H}_0^c}\ \delta[I_\tau-I^\prime]\Bigg\}.
  \end{split}
\end{equation}
\end{widetext}
Note that this equation is in parallel with our early quantum result in Eq.~(\ref{variance2}).

In the same manner,  the ensemble average $\langle e^{-2\beta \widetilde {W}} \rangle$ associated with an adiabatic process is found to be
\begin{widetext}
\begin{equation}\label{Cvariance4}
  \begin{split}
    \langle e^{-2\beta \widetilde{W}} \rangle & = \int_{I^{L}}^{I^{R}}[dI_0]_ {{H}_0^c}\ e^{2\beta H_0^c(I_0)}\rho_0(I_0) \int d\theta_0 \int d\theta\ \delta [\widetilde{\theta_\tau}-\theta ]  \\
    & \quad\Bigg\{ e^{-2\beta H^c_\tau(I^{R})}-\int_{I^{L}}^{I^{R}} d\left[e^{-2\beta H^c_\tau(I)}\right]_{{H}_0^c} \int_{I^{L}}^{I}[dI^\prime]_{{H}_0^c}\ \delta[I_0-I^\prime]\Bigg\},
  \end{split}
  \end{equation}
  where $\widetilde{W}$, $\widetilde{I_\tau}=I_0$, and $\widetilde\theta_\tau$ represent the values of
  work and action-angle variables at $t=\tau$ in an adiabatic process.  Needless to say, this result is analogous to Eq.~(\ref{variance3}) in the quantum case.

The difference in the square variance of $e^{-\beta W}$ between a general work protocol and an adiabatic process is then
given by
\begin{equation}\label{Cdifference}
   \langle e^{-2\beta W} \rangle - \langle e^{-2\beta \widetilde{W}} \rangle = \int_{I^{L}}^{I^R} d[-e^{-2\beta H^c_\tau(I)}]_ {H_0^{c}}\  \Lambda (I),
\end{equation}
where
\begin{equation}\label{CLambda}
\begin{split}
  \Lambda (I)= & \int_{I^L}^{I^{R}}[dI_0]_ {H_0^c}\ e^{2\beta H_0^c(I_0)}\rho_0(I_0)\int d\theta_0\ \Bigg\{\int d\theta\ \delta(\theta_\tau-\theta) \int_{I^L}^{I}[dI^\prime]_ {{H}_0^c}\ \delta(I_\tau-I^\prime) \\
    &  -\int d\theta\ \delta(\widetilde{\theta}_\tau-\theta) \int_{I^{L}}^{I} [dI^\prime]_ {{H}_0^c}\ \delta(I_0-I^\prime) \Bigg\}.
\end{split}
\end{equation}
\end{widetext}
 Equations (\ref{Cdifference}) and (\ref{CLambda}) are classical analogs of Eqs.~(\ref{difference1}) and (\ref{Xi}).
With the assumption that $H_0^c$ and $H_{\tau}^c$ can be connected by an adiabatic process, the instantaneous frequency $\omega_t = \frac{\partial H^c_t(I_t)}{\partial I_t}$ cannot be zero during the entire work protocol with $0\leq t \leq \tau$.
 Therefore, if for $I(n)= n\Delta I + I^L$, $H_0^c[I(n+1)]>H_0^c[I(n)]$, then $H_\tau^c[(I(n+1)]>H_\tau^c[(I(n)]$, namely, the ordering of the energy in terms of the value of the action variable is the same for $H_0^c$ and $H^c_\tau$.
  Imagine this is not the case, then
  at a particular time $0<t<\tau$, we must have $H_t^c[(I(n+1)]=H_t^c[(I(n)]$ and hence $\omega_t=0$.
  With these insights, it is now clear that the factor
  $d[-e^{-2\beta H^c_\tau(I)}]_ {H_0^{c}}$ in Eq.~(\ref{Cdifference}) cannot be negative.

One can split the integration in Eq.~(\ref{CLambda}) over $[dI_0]_{H_0^c}$ into two intervals: from $I^L$ to $I$ and from $I$ to $I^R$. This leads us to
%Combining Eqs.~(\ref{Cnormalize}), (\ref{Cspectrum}) and (\ref{Cdensityorder}), Eq.~(\ref{CLambda}) becomes
\begin{widetext}
\begin{equation}\label{CLambda2}
  \begin{split}
    \Lambda(I) & =
       \int_{I_{L}}^{I}[dI_0]_{{H}_0^c}\ \frac{e^{\beta H_0^c(I_0)}}{Z_0^c}\int d\theta_0\ \Bigg\{-\int d\theta\ \delta(\widetilde{\theta}_\tau-\theta) +\int d\theta\ \delta(\theta_\tau-\theta)\int_{I^L}^{I}[dI^\prime]_{{H}_0^c} \ \delta(I_\tau-I^\prime)\Bigg\}\\
      & \quad +\ \int_I^{I^{R}}[dI_0]_{{H}_0^c}\  \frac{e^{\beta H_0^c(I_0)}}{Z_0^c}\int  d\theta_0 \int d\theta\ \delta(\theta_\tau-\theta) \int_{I^L}^{I} [dI^\prime]_{{H}_0^c}\ \delta(I_\tau-I^\prime)\\
  \end{split}
  \end{equation}
 Further noticing that (i) the expression inside the $\{\cdot\}$ in Eq.~(\ref{CLambda2}) is not positive, (ii) the integrals over the $\delta$ functions in the second line of Eq.~(\ref{CLambda2}) is not negative, we immediately have the following
 \begin{equation} \label{CLambda3}
 \begin{split}
 \Lambda(I) & \geq \frac{e^{\beta H_0^c(I)}}{Z_0^c} \int_{I^{L}}^{I}[dI_0]_{{H}_0^c}\int d\theta_0 \Bigg\{-\int d\theta\ \delta(\widetilde{\theta}_\tau-\theta) +\int d\theta\ \delta(\theta_\tau-\theta) \int_{I^{L}}^{I} [dI^\prime]_{{H}_0^c}\ \delta(I_\tau-I^\prime)\Bigg\}\\
      & \quad +\ \frac{e^{\beta H_0^c(I)}}{Z_0^c} \int_I^{I^{R}}[dI_0]_{{H}_0^c}\int d\theta_0 \int d\theta \ \delta(\theta_\tau-\theta) \int_{I^L}^{I}[dI^\prime]_{{H}_0^c}\ \delta(I_\tau-I^\prime)\\
      & =  -\frac{e^{\beta H_0^c(I)}}{Z_0^c} \int_{I^L}^{I}[dI_0]_{{H}_0^c}\int d\theta_0 \int d\theta\ \delta(\widetilde{\theta}_\tau-\theta)  \\
      & \quad +\ \frac{e^{\beta H_0^c(I)}}{Z_0} \int_{I^L}^{I^R}[dI_0]_{{H}_0^c}\int d\theta_0 \int d\theta\ \delta(\theta_\tau-\theta) \int_{I^L}^{I}[dI^\prime]_{{H}_0^c}\ \delta(I_\tau-I^\prime)\\
      & = -\frac{e^{\beta H_0^c(I)}}{Z_0^c} \int_{I^L}^{I}[dI_0]_{{H}_0^c}\int d\theta_0   +\ \frac{e^{\beta H_0^c(I)}}{Z_0^c} \int d\theta  \int_{I^L}^{I}[dI^\prime]_{{H}_0^c}\\
       &=0 .
  \end{split}
\end{equation}
\end{widetext}
 In reaching the last two steps, we have used the fact that $\int_{I^L}^{I^R}[dI_0]_{{H}_0^c}\int d\theta_0\ \delta(\theta_\tau-\theta) \delta(I_\tau-I^\prime)=1$.   Hence, we have proved that for a classical adiabatic process with the above-assumed conditions (analogous to the no-level-crossing assumption in the quantum case),
\begin{equation}
\sigma^2 \langle e^{-\beta W} \rangle \geq \sigma^2 \langle e^{-\beta \widetilde{W}}\rangle,
\end{equation}
the principle of minimal work fluctuations in the classical domain.

Before ending this section, we remark that our technique here can be adapted to prove a classical version of the minimal (quantum) work principle obtained in Ref.~\cite{Allahverdyan.05.PRE}.  Considering the importance of the minimal work principle and the relevance of classical statistics in nanoscale systems with relatively high temperature,  in Appendix B we indeed give such a proof, which is similar to our ideas here proving the principle of minimal work fluctuations.
%%%%%%%%%%%%%%%%%%%%%%%%%%%%%%%%%%%%%%%%%%%%%%%%%%%%%%%%%%%%%%%%%%%%%%%%%%%%%%%%%%%%%
\section{Conclusion}
We have obtained a general principle regarding the minimal work fluctuations for thermally isolated systems initially prepared at equilibrium and then subject to a work protocol.  Specifically, if the initial and final states can be connected by an adiabatic process as stated in the seminal adiabatic theorem, then the variance of $e^{-\beta W}$ reaches the lower bound, as compared with all other processes operating between the same initial and final system Hamiltonians.  This is true in both the quantum and classical domains.  It is now clear  that an adiabatic process yields not only the minimal average work, but also the minimal fluctuations in $e^{-\beta W}$. This main result represents a somewhat counter-intuitive but fundamental understanding of work fluctuations.
The actual proof of our principle of minimal work fluctuations also indicates that the lower bound of work fluctuations can be equally reached by assisted adiabatic processes, such as those realized by STA.  Therefore, reaching the minimal fluctuations in work does not necessarily require a work protocol to be slow.  In addition to providing new insights into fluctuations phenomena in small systems, the results of this work should be of interest to the design of reliable and efficient energy devices at nano and micro scales.
%%%%%%%%%%%%%%%%%%%%%%%%%%%%%%%%%%%%%%%%%%%%%%%%%%%%%%%%%%%%%%%%%%%%%%%%%%%%%

\appendix
\section{Ensemble-based quantum optimal control theory}
 The quantum optimal control theory (OCT) we considered in order to suppress the fluctuations in $e^{-\beta W}$ is somewhat different from a traditional case due to two aspects. First, we need to handle a thermal ensemble~(see Eq.~(\ref{density1}) instead of a single quantum state as the initial state. Second, the quantity we need to optimize is based on two-time energy measurements, which is not an observable. Because of these peculiarities,  it is necessary to outline some details in our OCT calculations.

First, to suppress the fluctuations in $e^{-\beta W}$, an additional control field is considered, with the total Hamiltonian $H$ of the system given by
\begin{equation}\label{Htotal}
  H(t)=H_0(t)+H^{\text{OCT}}(A(t)),
\end{equation}
where $H^{\text{OCT}}$ is the control Hamiltonian and $A(t)$ is the time-dependent amplitude of a control field. The time-evolving  state $|\phi_i\rangle$ obeys the sch\"{o}rdinger equation:
\begin{equation}\label{schordinger}
  |\dot{\phi}_i(t)\rangle=-\text{i}H|\phi_i(t)\rangle,
\end{equation}
where $|\phi_i\rangle$ denotes the $i$-th state in the initial thermal ensemble in the energy basis.

Consider next a certain quantity described by Eq.~(\ref{average}) as a target control function,  denoted as $L_1$.  The job is to minimize the following:
\begin{equation}\label{L1}
  L_1=\sum_{i,j=1}^N p_i |\langle \psi_j^\prime|\phi_i(\tau)\rangle|^2 f(E_i,E_j^\prime),
\end{equation}
where $|\phi_i(\tau)=U|\psi_i\rangle$ is the final state evolved from the initial state $|\psi_i\rangle$ (eigenstate of $H_0$).
For a control problem, typically a cost function is also needed to reflect a cost-related constraint. This cost function can be constructed as
\begin{equation}\label{L2}
  L_2=\frac{1}{2}\int_0^\tau \kappa A^2(t) dt,
\end{equation}
where $A(t)$ is above-mentioned amplitude of the control field and $\kappa$ is a weightage factor. The overall target function can then be defined as $J=L_1+L_2$. That is, the problem is now to minimize $J$ under the general dynamical constraint reflected by the Schr\"{o}dinger equation.

To proceed we introduce $N$ Lagrange multiplier vectors as a function of $t$, denoted by $\{|l_i(t)\rangle, i\in N \}$. We then minimize $\bar{J}$ instead, with
\begin{equation}\label{Jbar}
  \bar{J}=L_1+L_2+\sum_{i=1}^{N}\int_0^{\tau} [\langle l_i(t)| \dot{\phi}_i \rangle+\text{i}\langle l_i| H| \phi_i(t) \rangle] dt.
\end{equation}
Let $|\delta \phi_i(t)\rangle$ be the variation in $|\phi_i(t)\rangle$ due to $\delta A(t)$, an arbitrary variation in $A(t)$, then the variation in $\bar{J}$ due to $\delta A(t)$ is found to be
\begin{equation}\label{deltaJbar}
  \begin{split}
    \delta \bar{J} = & \sum_{i,j=1}^N p_i [\langle \psi_j^\prime | \delta \phi_i(\tau) \rangle +h.c.]f(E_i,E_j^\prime) \\
      & +\sum_{i=1}^N [\langle l_i(\tau)|\delta \phi_i(\tau) \rangle+h.c.] \\
      & -\sum_{i=1}^N \int_0^{\tau} [\langle \dot{l}_i(t)|\delta \phi_i(t) \rangle+h.c.]dt \\
      & +\sum_{i=1}^N \int_0^{\tau} \text{i}[\langle l_i(t)|H|\delta \phi_i(t \rangle)-h.c.]dt\\
      & +\sum_{i=1}^N \int_0^{\tau} \text{i}[\langle l_i(t)|\frac{\partial H}{\partial A}| \phi_i(t)\rangle-h.c.]\delta A dt \\
      & +\int_0^{\tau} \kappa A(t) \delta A dt.
  \end{split}
\end{equation}
To minimize $\bar{J}$ we let $\delta \bar{J}=0$. Since the variation is arbitrary, one has the following relations:
\begin{equation}\label{relation}
  \left\{
     \begin{split}
         & |\dot{l}_i(t)\rangle = -\text{i}H|l_i(t) \rangle\\
         & 2\textrm{Im}[\langle l_i(t)|\frac{\partial H}{\partial A}| \phi_i(t)\rangle]+\kappa A(t)=0 \\
         & p_i\sum_{j=1}^N |\psi_j^\prime\rangle f(E_i,E_j^\prime)+|l_i(\tau)\rangle=0.
     \end{split}
  \right.
\end{equation}
The above list of relations can be numerically solved by an iteration procedure \cite{Shi.90.JCP}.  In our actual calculations, in precisely the same manner as Ref.~\cite{Xiao.14.PRE}, we introduced a time dependence to the ``penalty factor" $\kappa$ to ensure that the OCT control field is zero at $t=0$ and at $t=\tau$.
\section{A classical version of the minimal work principle}
An extension of the minimal work principle proven in Ref.~\cite{Allahverdyan.05.PRE} to the classical domain can be carried out, following essentially the same steps used in our above proof of the minimal classical work fluctuations.   With the same notation as in the main text,   the average work for a general process starting from $H_0^{c}$ and ending with $H_\tau^c$ is
\begin{widetext}
\begin{equation}\label{Cvariance2-A}
  \langle {W} \rangle +\langle H_0^c\rangle   = \int_{I^{L}}^{I^{R}} [dI_0]_{{H}_0^c}\ \rho_0(I_0)\int d\theta_0 \int_{I^{L}}^{I^{R}}[dI]_{{H}^c_0}\ H^c_\tau(I) \int d\theta\ \delta [I_\tau-I] \delta [\theta_\tau-\theta ].
\end{equation}
Upon integration parts, we have
\begin{equation}\label{Cvariance3-A}
  \begin{split}
    \langle  W \rangle + \langle H_0^c\rangle & = \int_{I^{L}}^{I^{R}}[dI_0]_ {{H}_0^c}\ \rho_0(I_0) \int d\theta_0 \int d\theta\ \delta [\theta_\tau-\theta ]  \\
    & \quad\Bigg\{ H^c_\tau(I^{R})-\int_{I^{L}}^{I^{R}} d\left[ H^c_\tau(I)\right]_{{H}_0^c} \int_{I^{L}}^{I}[dI^\prime]_{{H}_0^c}\ \delta[I_\tau-I^\prime]\Bigg\}.
  \end{split}
\end{equation}
The parallel result for an adiabatic process is
\begin{equation}\label{Cvariance4-A}
  \begin{split}
    \langle \widetilde{W}
     \rangle + \langle H_0^c\rangle   & = \int_{I^{L}}^{I^{R}}[dI_0]_ {{H}_0^c}\ \rho_0(I_0) \int d\theta_0 \int d\theta\ \delta[\widetilde{\theta}_\tau-\theta ]  \\
    & \quad\Bigg\{ H^c_\tau(I^{R})-\int_{I^{L}}^{I^{R}} d\left[ H^c_\tau(I)\right]_{{H}_0^c} \int_{I^{L}}^{I}[dI^\prime]_{{H}_0^c}\ \delta[I_0-I^\prime]\Bigg\}.
  \end{split}
\end{equation}
One hence finds the difference between $\langle W\rangle$ and $\langle \widetilde{W}\rangle$:
\begin{equation}\label{Cdifference-A}
   \langle  {W} \rangle - \langle \widetilde{W}\rangle = \int_{I^{L}}^{I^R} d[ H^c_\tau(I)]_{H_0^{c}}\  \Theta (I),
\end{equation}
where
\begin{equation}\label{CLambda-A}
\begin{split}
  \Theta (I)= &  \int_{I^L}^{I^{R}}[dI_0]_ {H_0^c}\ \rho_0(I_0)\int d\theta_0\ \Bigg\{-\int d\theta\ \delta(\theta_\tau-\theta) \int_{I^L}^{I}[dI^\prime]_ {{H}_0^c}\ \delta(I_\tau-I^\prime) \\
    &  +\int d\theta\ \delta(\widetilde{\theta}_\tau-\theta) \int_{I^{L}}^{I} [dI^\prime]_ {{H}_0^c} \delta(I_0-I^\prime) \Bigg\}.
\end{split}
\end{equation}
Note that due to the same-ordering condition,  $ d[H^c_\tau(I)]_{H_0^{c}}$ is not negative.
Splitting the integration in Eq.~(\ref{CLambda-A}) over $[dI_0]_{H_0^c}$ into two intervals: from $I^L$ to $I$ and from $I$ to $I^R$, we have
%Combining Eqs.~(\ref{Cnormalize}), (\ref{Cspectrum}) and (\ref{Cdensityorder}), Eq.~(\ref{CLambda}) becomes
\begin{equation}\label{CLambda2-A}
  \begin{split}
    \Theta(I) & =
       \int_{I_{L}}^{I}[dI_0]_{{H}_0^c}\ \frac{e^{-\beta H_0^c(I_0)}}{Z_0^c} \int d\theta_0\ \Bigg\{ \int d\theta\ \delta(\widetilde{\theta}_\tau-\theta) -\int d\theta\ \delta(\theta_\tau-\theta)\int_{I^L}^{I}[dI^\prime]_{{H}_0^c} \ \delta(I_\tau-I^\prime)\Bigg\}\\
      & \quad -\ \int_I^{I^{R}}[dI_0]_{{H}_0^c}\  \frac{e^{-\beta H_0^c(I_0)}}{Z_0^c} \int  d\theta_0 \int d\theta\ \delta(\theta_\tau-\theta) \int_{I^L}^{I} [dI^\prime]_{{H}_0^c}\ \delta(I_\tau-I^\prime)\\
  \end{split}
  \end{equation}
  For similar reasons as we derive Eq.~(\ref{CLambda3}), we now have the following inequality,
  \begin{equation}
  \begin{split}
\Theta(I) & \geq  \frac{ e^{-\beta H_0^c(I)}}{Z_0^c} \int_{I^{L}}^{I}[dI_0]_{{H}_0^c}\int d\theta_0 \Bigg\{\int d\theta\ \delta(\widetilde{\theta}_\tau-\theta) -\int d\theta\ \delta(\theta_\tau-\theta) \int_{I^{L}}^{I} [dI^\prime]_{{H}_0^c}\ \delta(I_\tau-I^\prime)\Bigg\}\\
      & \quad -\frac{e^{-\beta H_0^c(I)}}{Z_0^c} \int_I^{I^{R}}[dI_0]_{{H}_0^c}\int d\theta_0 \int d\theta \ \delta(\theta_\tau-\theta) \int_{I^L}^{I}[dI^\prime]_{{H}_0^c}\ \delta(I_\tau-I^\prime)\\
      & =  \frac{e^{-\beta H_0^c(I)}}{Z_0^c} \int_{I^L}^{I}[dI_0]_{{H}_0^c}\int d\theta_0 \int d\theta\ \delta(\widetilde{\theta}_\tau-\theta)  \\
      & \quad -\ \frac{e^{-\beta H_0^c(I)}}{Z_0} \int_{I^L}^{I^R}[dI_0]_{{H}_0^c}\int d\theta_0 \int d\theta\ \delta(\theta_\tau-\theta) \int_{I^L}^{I}[dI^\prime]_{{H}_0^c}\ \delta(I_\tau-I^\prime)\\
      & = \frac{e^{-\beta H_0^c(I)}}{Z_0^c} \int_{I^L}^{I}[dI_0]_{{H}_0^c}\int d\theta_0  - \frac{e^{-\beta H_0^c(I)}}{Z_0^c} \int d\theta  \int_{I^L}^{I}[dI^\prime]_{{H}_0^c}\\
       &=0 .
\end{split}
\end{equation}
\end{widetext}
This finally leads us to the conclusion that $\langle {W}\rangle\geq \langle \widetilde{W}\rangle$.

%\section{Reference}
%\bibliography{All}{}

\newpage
\begin{thebibliography}{99}%
\bibitem{Jarzynski.97.PRL} C. Jarzynski, Phys. Rev. Lett. {\bf 78}, 2690 (1997).
\bibitem{Jarzynski.97.PRE} C. Jarzynski, Phys. Rev. E {\bf 56}, 5018 (1997).
\bibitem{Mukamel.03.PRL} S. Mukamel, Phys. Rev. Lett. {\bf 90}, 170604 (2003).
\bibitem{Tasaki.00.apc} H. Tasaki, arXiv preprint cond-mat/0009244 (2000).
\bibitem{hanggireview} M. Campisi, P. H¨anggi, and P. Talkner, Rev. Mod. Phys. {\bf 83}, 771
(2011).
\bibitem{Crooks.99.PRE} G. E. Crooks, Phys. Rev. E {\bf 60}, 2721 (1999).
\bibitem{Xiao.14.PRE} G.~Y.~Xiao and J.~B.~Gong, Phys. Rev. E {\bf 90}, 052132 (2014).
\bibitem{Deng.13.PRE} J.~W.~Deng, Q.-h. Wang, Z.~H.~ Liu, P.~H\"{a}nggi, and J.~B. Gong, Phys. Rev. E {\bf 88}, 062122 (2013).
\bibitem{Abah.12.PRL} O. Abah, J. Ro{\ss}nagel, G. Jacob, S. Deffner, F. Schmidt-Kaler, K. Singer, and E. Lutz, Phys. Rev. Lett. {\bf 109}, 203006 (2012).
\bibitem{Bergenfeldt.14.PRL} C. Bergenfeldt, P. Samuelsson, B. Sothmann, C. Flindt, and M. B\"{u}ttiker, Phys. Rev. Lett. {\bf 112}, 076803 (2014).
\bibitem{Zhang.14.PRL} K. Zhang, F. Bariani, and P. Meystre, Phys. Rev. Lett. {\bf 112}, 150602 (2014).
\bibitem{Zheng.14.PRE} Y. Zheng and D. Poletti, Phys. Rev. E {\bf 90}, 012145 (2014).
\bibitem{Zheng.15.a} Y. Zheng and D. Poletti, arXiv:1504.02183 (2015).
\bibitem{gaoyang-preprint} G.~Y.~Xiao and J.~B.~Gong, arXiv.1503.00784.
\bibitem{Bender.00.JPAMG} C. M. Bender, D. C. Brody, and B. K. Meister, J. Phys. A: Math. Gen. {\bf 33}, 4427 (2000).
\bibitem{Quan.07.PRE} H. T. Quan, Y.-x. Liu, C. P. Sun, and F. Nori, Phys. Rev. E {\bf 76}, 031105 (2007).
\bibitem{adiabatic}M. Born and V. A. Fock, ``Beweis des Adiabatensatzes,'' Z. Phys. A {\bf 51} 165 (1928).

\bibitem{LZ1}L.~D.~Landau, Phys. Z. {\it Sowjetunion} {\bf 2} 46 (1932).
\bibitem{LZ2} C.~Zener, Proc. R. Soc. {\bf A 137}, 696 (1932).
\bibitem{Gelbwaser-Klimovsky.15.a} D. Gelbwaser-Klimovsky, W. Niedenzu, and G. Kurizki, arXiv preprint arXiv:1503.01195 (2015).

\bibitem{Allahverdyan.05.PRE} A. E. Allahverdyan and T. M. Nieuwenhuizen, Phys. Rev. E {\bf 71}, 046107 (2005).
\bibitem{Rice} M.~Demirplak and S.~A.~Rice, J.~Phys.~Chem. A{\bf 107}, 9937 (2003); J. Phys. Chem. B {\bf 109}, 6838 (2005).
\bibitem{Berry} M.~V.~Berry, J.~Phys.~A: Maths.~Theor.~{\bf 42}, 365303 (2009).
\bibitem{Torrontegui.13.AAMOP} E. Torrontegui, S. Ib\'{a}\~{n}ez, S. Mart\'{\i}nez-Garaot, M. Modugno, A. el Campo, D. Gu\'{e}ry-Odelin, A. Ruschhaupt, X. Chen, and J. G. Muga, Adv. At. Mol. Opt.~Phys. {\bf 62}, 117 (2013).
\bibitem{Chen.10.PRL} X. Chen, I. Lizuain, A. Ruschhaupt, D. Gu\'{e}ry-Odelin, and J. G. Muga, Phys. Rev. Lett. {\bf 105}, 123003 (2010).
\bibitem{Ibanez.12.PRL} S. Ib\'{a}\~{n}ez, X. Chen, E. Torrontegui, J. G. Muga, and A. Ruschhaupt, Phys. Rev. Lett. {\bf 109}, 100403 (2012).
 \bibitem{Campo.13.PRL} A. del Campo, Phys. Rev. Lett. {\bf 111}, 100502 (2013).


\bibitem{Peirce.88.PRA} A. P. Peirce, M. A. Dahleh, and H. Rabitz, Phys. Rev. A {\bf 37}, 4950 (1988).
\bibitem{Shi.90.JCP} S. Shi and H. Rabitz, J. Chem. Phys. {\bf 92}, 364 (1990).
\bibitem{Shi.91.CPC} S. Shi and H. Rabitz, Comput. Phys. Commun. {\bf 63}, 71 (1991).
\bibitem{Judson.92.PRL} R. S. Judson and H. Rabitz, Phys. Rev. Lett. {\bf 68}, 1500 (1992).

\bibitem{Riviello.14.PRA} G. Riviello, C. Brif, R. Long, R.-B. Wu, K. M. Tibbetts, T.-S. Ho, and H. Rabitz, Phys. Rev. A {\bf 90}, 013404 (2014).
\bibitem{Jarzynski.07.CRP} C. Jarzynski, C. R. Phys. {\bf 8}, 495 (2007).

\end{thebibliography}

\end{document}